\documentclass[lettersize,journal]{IEEEtran}
\usepackage{array}
\usepackage{amsfonts}
\usepackage{amsbsy}
\usepackage{amssymb}
\usepackage{times}
\usepackage{graphicx}
\usepackage{enumerate}
\usepackage[usenames]{color}
\usepackage[dvips]{pstcol}
\usepackage{epstopdf}
\usepackage[caption=false,font=normalsize,labelfont=sf,textfont=sf]{subfig}
\usepackage{cite}
\usepackage{amssymb}
\usepackage{amsfonts}
\usepackage{graphicx}
\usepackage{epsfig}
\usepackage{psfrag}
\usepackage{xcolor}
\usepackage{amsfonts, bm}
\usepackage{epstopdf}
\usepackage{cite}
\usepackage{color}
\usepackage{xcolor}
\usepackage{subfig}
\usepackage{verbatim}
\usepackage{multirow}
\usepackage{array}
\usepackage{booktabs}
\usepackage{amsthm}
\usepackage{makecell}

\usepackage[linesnumbered, ruled]{algorithm2e}
\usepackage{algpseudocode}
\usepackage{amsmath}

\IEEEoverridecommandlockouts

\begin{document}

	\title{Alleviating Distortion Accumulation in Multi-Hop Semantic Communication}
\author{Guangyi Zhang, \textit{Student Member, IEEE}, Qiyu Hu, \textit{Student Member, IEEE}, Yunlong Cai, \textit{Senior \\Member, IEEE}, and Guanding Yu, \textit{Senior Member, IEEE}
	\thanks{ G. Zhang, Q. Hu, Y. Cai, and G. Yu are with the College of Information Science and Electronic Engineering, Zhejiang University, Hangzhou 310027, China (e-mail: zhangguangyi@zju.edu.cn; qiyhu@zju.edu.cn; ylcai@zju.edu.cn; yuguanding@zju.edu.cn). 
} }
	\maketitle
	\vspace{-3.3em}
	\begin{abstract}
		Recently, semantic communication has  been investigated to boost the performance of end-to-end image transmission systems. However, existing semantic approaches are generally based on deep learning and belong to lossy transmission. Consequently, as the receiver continues to transmit received images to another device, the distortion of images accumulates with each transmission. Unfortunately, most recent advances overlook this issue and only consider single-hop scenarios, where images are transmitted only once from a transmitter to a receiver. In this letter, we propose a novel framework of a multi-hop semantic communication system. To address the problem of distortion accumulation, we introduce a novel recursive training method for the encoder and decoder of semantic communication systems. Specifically, the received images are recursively input into the encoder and decoder to retrain the semantic communication system. This empowers the system to handle distorted received images and achieve higher performance. Our extensive simulation results  demonstrate  that the proposed methods significantly alleviate distortion accumulation in multi-hop semantic communication.
	\end{abstract}
	
	
	\IEEEpeerreviewmaketitle
	
	\section{Introduction}
	In current wireless communication systems, data transmission commonly involves a two-step encoding process: source coding and channel coding \cite{Wenyu_TCOM,Bouurt_TCCN,Shuaishuai_CL}. Source coding techniques, such as joint photographic experts group (JPEG) and better portable graphics (BPG), are typically used for image coding. Additionally, error-correcting channel codes, such as low-density parity check (LDPC) codes, are employed to enhance transmission reliability. Despite the benefits of these separate coding techniques in improving communication efficiency, they may not fully support data transmission in future networks \cite{Qiyu_TWC}. The rapid growth of intelligent services, such as Virtual Reality (VR) and the Internet of Everything (IoE), calls for further improvements in communication efficiency \cite{Wangting_CST, Bingxuan_arxiv}. In this context, semantic communication, which focuses on semantic exchange rather than solely minimizing the bit error rate, has gained significant attention.
	
	By implementing source coding and channel coding jointly based on deep neural networks (DNNs), semantic communication has shown its exciting  capabilities  in various data transmission tasks \cite{Jialong_TCVT}. In particular, the joint source channel coding (JSCC)-based semantic method proposed in \cite{Jialong_TCVT} employs the attention mechanism to adjust the extracted features under different channel signal-to-noise ratio (SNR) conditions, significantly improving  the adaptability of semantic communication to various channel conditions. In \cite{Haotian_arxiv}, the authors conceived a novel paradigm for image transmission using feedback from the receiver. As introduced in \cite{Sixian_JSAC}, a new class of semantic communication methods has been developed to achieve end-to-end single-hop semantic transmission using non-linear transform coding.

\begin{figure*}[t]
	\begin{centering}
		\includegraphics[width=0.72\textwidth]{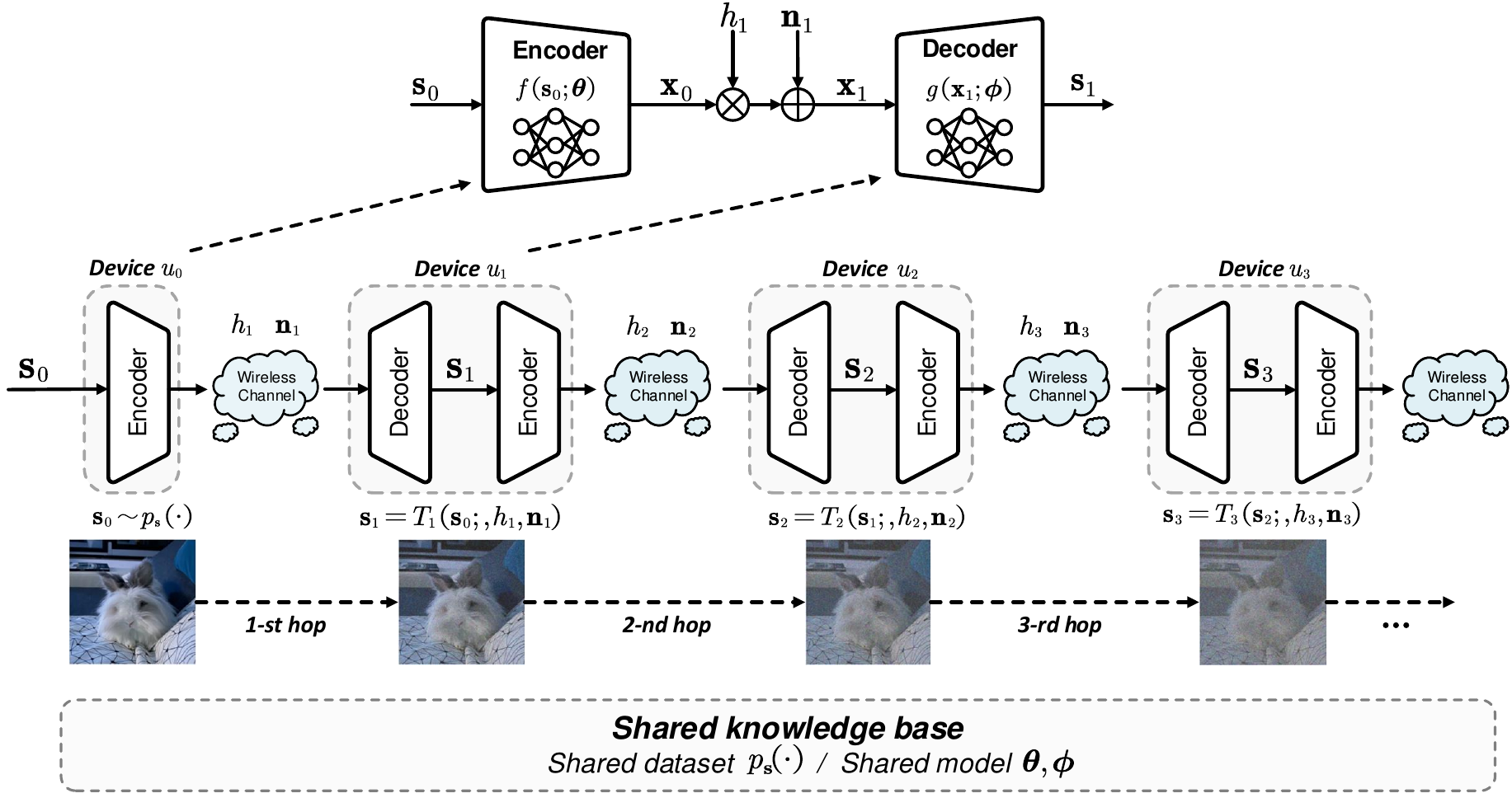}
		\par\end{centering}
	\caption{The framework of proposed multi-hop semantic communication.}
	\label{ProbModel}
	\end{figure*}

	In comparison with  separate coding techniques, semantic communication systems are capable of achieving better performance, especially at low SNR regimes, and are known for overcoming the cliff effect \cite{Sixian_JSAC, Jianhao_IOTJ, Han_CL}. That is, when channel conditions deteriorate beyond a certain threshold, it will be difficult for the decoder to effectively recover the transmitted information, resulting in a cliff edge degradation in system performance.
	Nonetheless, existing works primarily focus on single-hop communication systems. It is also important to consider the multi-hop semantic communication, which are basically used to increase the coverage of network for cases that the basement is not in direct line of sight of the signal \cite{Ming-Chun_TCOM}. Unfortunately, as a paradigm of lossy transmission, semantic communication faces a new challenge of distortion accumulation in multi-hop scenarios, which does not exist in separate coding systems. In particular, the distortion of signal will accumulate with the number of transmissions, resulting in a gradual increase in noise and a decrease in image's quality, as illustrated in Fig. \ref{ProbModel}.

	There are two main factors that contribute to the problem of distortion accumulation. On the one hand, there will be a distortion between the transmitted image and the received image at each transmission, since semantic communication systems are generally lossy systems. In this case, if the receiver  requires to transmit the received image to another device, as shown in Fig. \ref{ProbModel}, the receiver will perform as a new transmitter, and its encoder needs to handle the distorted images. However, this is difficult to achieve since the distorted images generally have different statistical characteristics from that of the source image, while the encoders and decoders in different devices are trained with clean source images \cite{Huiqiang_WCL, Hongwei_JSAC,Mingyu_TCCN}. On the other hand, the next hop of transmission also introduces perturbations, such as additive white Gaussian noise (AWGN). As a result, the distortion will accumulate over the number of transmissions. To the best of our knowledge, existing studies mostly focus on single-hop point-to-point communication, and no prior work has specifically addressed the problem of distortion accumulation.
	
	In this letter, we propose a novel framework for a multi-hop semantic communication system with multiple devices, and develop a shared vision transformer-based semantic communication system (ViTSC) for each device to enable wireless image transmission. We model  the problem of distortion accumulation in multi-hop semantic communication. To address this issue, we develop a novel recursive training method with incremental weight scheme to train the encoder and decoder of ViTSC. Specifically, the training process of ViTSC is divided into multiple stages, and at each stage, the received images will be recursively fed into ViTSC to train the semantic communication system. Additionally, we introduce an incremental weighting scheme to assign different weights to the loss function at different stages, enabling it to handle received images with different degrees of distortion. Simulation results demonstrate that the proposed methods significantly alleviate distortion accumulation in multi-hop semantic communication.
	

	\section{Framework of Multi-hop Semantic Communication} \label{S2}
	In this section, we propose the framework of multi-hop semantic communication system for image transmission. To provide a comprehensive understanding, we begin by giving an overview of single-hop semantic communication and then proceed to introduce the proposed multi-hop semantic communication framework.

	\subsection{Single-Hop Semantic Communication}
	As shown in the upper part of Fig. \ref{ProbModel}, we consider a lossy wireless image transmission scenario, where device $u_0$ transmits an $n$-pixel image $\mathbf{s}_0 \in \mathbb{R}^n$ to the receiver. The image $\mathbf{s}_0$ is drawn from dataset $\mathcal{S}$ with probability $p_\mathbf{s}(\cdot)$. In particular, the transmitter employs an encoder $f_0(\cdot)$ to map $\mathbf{s}_0$ into a $k$-dimensional vector $\mathbf{x}_0 \in \mathbb{C}^k$, using an encoder parameterized by $\bm{\theta}$. Thus, the encoded feature vector can be represented by $\mathbf{x}_0=f_0(\mathbf{s}_0;\bm{\theta})$. Accordingly, the channel bandwidth ratio (CBR) is defined as $\rho = k/n$, denoting the number of channel uses per image pixel. Moreover, $\mathbf{x}_0$ is subject to a power constraint: $\frac{1}{k}\mathbb{E}\| \mathbf{x}_0\|^2 \leq 1$.

	Subsequently, the encoded vector $\mathbf{x}_0$ is directly transmitted over the wireless channel and received by the receiver. Specifically, the received vector is given by $\mathbf{x}_1=h_1\mathbf{x}_0+\mathbf{n}_1$, where $h_1 \in \mathbb{C}$ denotes the channel gain coefficient and $ \mathbf{n}_1 \sim \mathcal{CN}\left(0, \sigma^2_1 \mathbf{I}_k\right)$ is AWGN with average power $\sigma^2_1$. The channel transformation function is denoted as $w_1(\cdot)$. The receiver at device $u_1$ consisting of a decoder $g_1(\cdot)$ aims to reconstruct the corrupt signal $\mathbf{x}_1$ into $\mathbf{s}_1$, i.e., $\mathbf{s}_1 = g_1(\mathbf{x}_1;\bm{\phi})$, where $\bm{\phi}$ denotes the trainable parameters of the decoder and $\mathbf{s}_1$ can be regarded as an approximation of $\mathbf{s}_0$. Thus, the objective of a single-hop semantic communication is expressed as 
	\begin{equation} \label{o_sh}
		\min _{\bm{\theta},\bm{\phi}}  \mathbb{E}\left[  d(\mathbf{s}_0,g_1(\mathbf{x}_1;\bm{\phi})) \right],
	\end{equation}
	where $d(\cdot)$ denotes the distortion function.

	\subsection{Multi-Hop Semantic Communication}
	We propose a cooperative multi-hop semantic communication, as presented in Fig \ref{ProbModel}, which can be seen as series of multiple single-hop semantic communication system. In particular, it consists of a source device $u_0$ and $t$ target devices $u_1, u_2,...,u_t$. For clarity, we have set $t=3$  in Fig. \ref{ProbModel}.
	In the following, we employ subscripts to indicate the devices to which the vectors belong. For instance, $\mathbf{x}_0$ represents the source image in device $u_0$, and $\mathbf{s}_3$ represents the recovered image in device $u_3$. Moreover, we assume that all devices share a common empirical knowledge $\mathcal{S}$. This means they will have access to a shared dataset and utilize the encoder and decoder with shared parameters $\{\bm{\theta},\bm{\phi}\}$, trained based on the dataset $\mathcal{S}$.
	
	The data transmission process of the multi-hop semantic communication system is as follows. The source device $u_0$ transmits data $\mathbf{s}_0 \sim p_{\mathbf{s}}(\cdot)$ to device $u_1$, which then obtains $\mathbf{s}_1$. Subsequently, device $u_1$ becomes a new transmitter and further sends $\mathbf{s}_1$ to device $u_2$. This process continues until the data reaches its all intended destinations. Denote the transformation function from $\mathbf{s}_0$ to $\mathbf{s}_1$ as 
	\begin{equation}
		\mathbf{s}_1\triangleq T_1(\mathbf{s}_0; h_1, \mathbf{n}_1) = g_1(w_1(f_0(\mathbf{s}_0;\bm{\theta}); h_1,\mathbf{n}_1);\bm{\phi}).
	\end{equation}
	Accordingly, the transmission process of a multi-hop semantic communication system can be formulated as
	\begin{equation}
		\mathbf{s}_0 \xrightarrow{T_1\left(\cdot; h_1, \mathbf{n}_1\right)} \mathbf{s}_1 
		\xrightarrow{T_2\left(\cdot ; h_2, \mathbf{n}_2\right)} \mathbf{s}_2
		\xrightarrow{T_3\left(\cdot ; h_3, \mathbf{n}_3\right)} \mathbf{s}_3 \xrightarrow{\dots}	\mathbf{s}_t.
	\end{equation}
	Since the target of these devices should be to reconstruct the image as close as possible to $\mathbf{s}_0$, the objective for such multi-hop scenario becomes 
	\begin{equation}
		\min _{\{\bm{\theta},\bm{\phi}\}}  \mathbb{E}\left[  d(\mathbf{s}_0,\mathbf{s}_i) \right], \; i=1,2,...,t.
	\end{equation}

	\section{Recursive Training and Model Architecture} \label{S3}
	In this section, we model the problem of distortion accumulation in multi-hop semantic communication, and propose a recursive training algorithm to address the problem of distortion accumulation in multi-hop semantic communication. Moreover, we present the architecture of ViTSC for wireless image transmission.

	\begin{figure}[t]
		\begin{centering}
			\includegraphics[width=0.43\textwidth]{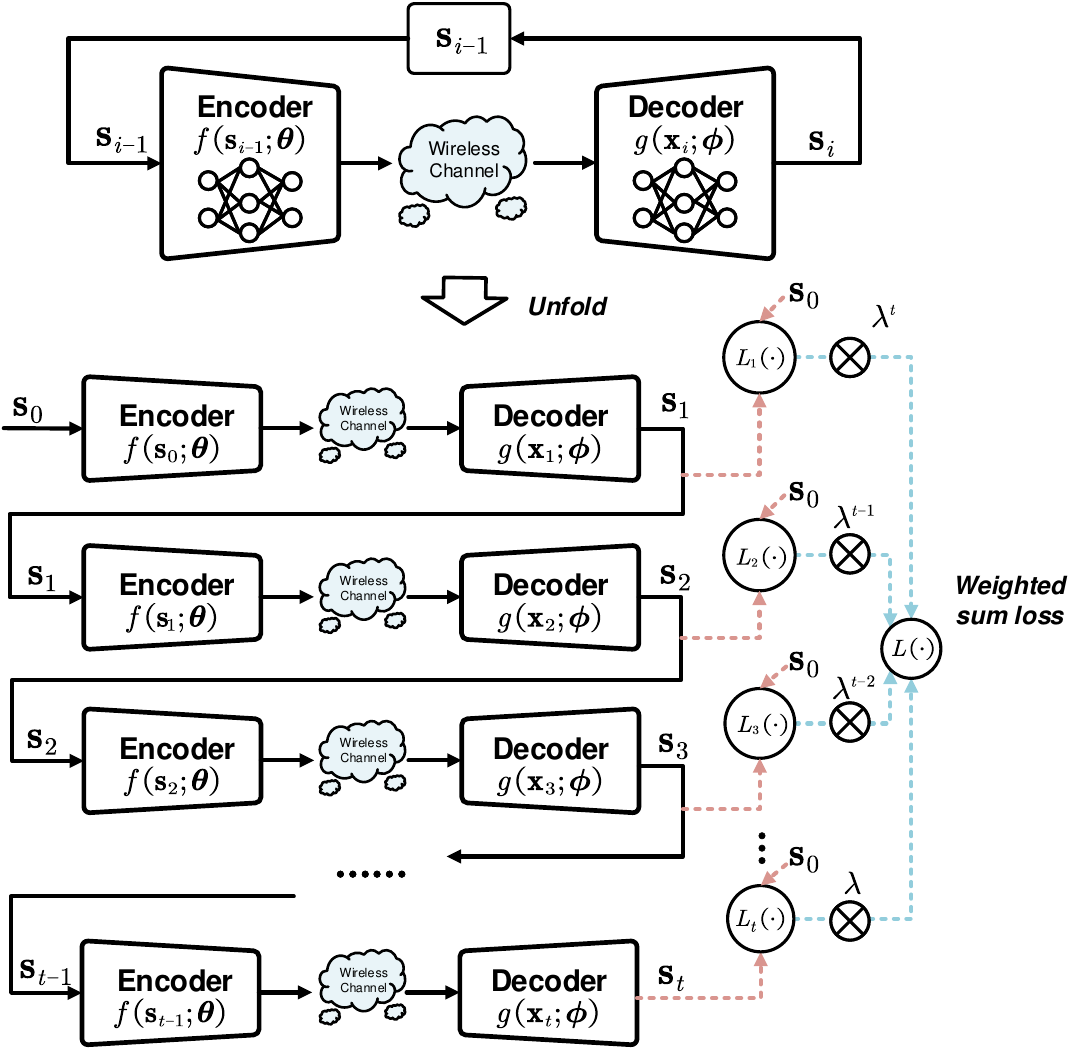}
			\par\end{centering}
		\caption{The illustrations of the recursive training method and the calculation of overall loss function.}
		\label{RecursiveTrain}
	\end{figure}
	\subsection{Distortion Accumulation}
	It is worth noting that the existing single-hop semantic communication schemes primarily focus on optimizing the end-to-end distortion between two devices, as given by (\ref{o_sh}). Intuitively, in a multi-hop scenario, following these existing solutions, each device will be equipped with the same encoder and decoder, which are trained with data from $\mathcal{S}$. However, such direct deployment will inevitably lead to the problem of distortion accumulation. The main reasons for this are as follows. 
	\begin{itemize}
		\item[(i)] Existing deep learning-based semantic communication systems are lossy transmission schemes, in which case the distortion will accumulate as the receiver continues to transmit received images to other devices.  As the images traverse through various devices, the cumulative effect of distortion can degrade the quality of the received images. 
		
		\item[(ii)] Both encoder and decoder are trained using data from $\mathcal{S}$, i.e., ${\bm{\theta},\bm{\phi}}\!=\!\arg \min_ {\bm{\theta},\bm{\phi}} \mathbb{E}_{\mathbf{s}_0 \sim p_{\mathbf{s}}}\left[ d(\mathbf{s}_0,g_1(\mathbf{x}_1;\bm{\phi})) \right]$. However, the statistical characteristics of the reconstructed images at other devices, such as $\mathbf{s}_i$ at device $u_i$, can significantly differ from $\mathbf{s}_0$. Consequently, it becomes challenging for device $u_i$ to encode $\mathbf{s}_i$ with an encoder trained on $\mathbf{s}_0 \sim p_{\mathbf{s}}(\cdot)$, resulting in performance degradation.
	\end{itemize}
	As a result, the images gradually become blurry as the number of transmissions increases, posing a significant challenge for the practical implementation of semantic communication models.

	\subsection{Recursive Training}

		\begin{algorithm}[t]
		\textbf{Input:} Training dataset $\mathcal{S}$, number of devices $t$, encoder and decoder with parameters $\{\bm{\theta},\bm{\phi}\}$ trained in single-hop scenario, and number of training epochs $M$. \\
		\textbf{Output:} Optimized parameters $\{\bm{\theta}^*\!,\!\bm{\phi}^{*}\}$. \\
		\BlankLine
		\For{$m = 1$ \KwTo $M$}{
			Sample  data $\mathbf{s}_0$ from $\mathcal{S}$.\\
			\For{$i = 1$ \KwTo $t$}{ 
				Sample $h_i$ and $\mathbf{n}_i$ based on their distributions.\\
				Caculate $\mathbf{s}_i=T_i\left(\mathbf{s}_{i-1}; h_i, \mathbf{n}_i\right)$.\\
				Calculate the $i$-th loss function $L_i\left(\mathbf{s}_{i-1}, \mathbf{s}, h_i, \mathbf{n}_i\right)$ based on (\ref{ele_loss}).\\
			}
			Compute the sum loss $L\left(\mathbf{s}_0, \boldsymbol{\theta}, \boldsymbol{\phi}\right)$ according to (\ref{sum_loss}).\\
			Upadate $\{\bm{\theta},\bm{\phi}\}$ using gradient $\{\nabla_{\!\bm{\theta}\!}L,\nabla_{\!\bm{\phi}\!}L\}$.
		}
		
		\caption{Recursive training algorithm}
		\label{RecursiveAlgorithm}
	\end{algorithm}
	To alleviate the distortion accumulation, we propose to incorporate the reconstructed images into the training process, enabling the encoder and decoder to learn to handle noisy reconstructed images.
	We consider a multi-hop scenario that has $t$ devices, our goal is to find a pair of encoder $f$ and decoder $g$ with parameters $\{\bm{\theta},\bm{\phi}\}$ that can support transmitting the received data at different devices. To achieve this, the output of the decoder is recursively input into the encoder to train the model, as illustrated in Fig. \ref{RecursiveTrain}. This process is equivalent to adding $\mathbf{s}_i$, $i=1,2,...,t$, to the training dataset. Specifically, the training process can be unfolded into $t$ phases, as shown in Fig. \ref{RecursiveTrain}, corresponding to $t$ transmissions in the multi-hop semantic communication system. At each phase, we calculate the distortion between the received image and $\mathbf{s}_0$ as the loss function. Then, we will obtain in total of $t$ loss functions. In particular, the $i$-th loss $L_i$ is defined as the distortion between the $i$-th input $\mathbf{s}_{i-1}$ and the $i$-th output $\mathbf{s}_i$:
	\begin{equation}\label{ele_loss}
		L_i (\mathbf{s}_{i-1},\mathbf{s}_i,h_i, \mathbf{n}_i) \triangleq d(\mathbf{s}_{i-1},\mathbf{s}_{i})=d(\mathbf{s}_{i-1},T_i(\mathbf{s}_{i-1},h_i,\mathbf{n}_i)),
	\end{equation}
	where we use mean squared error (MSE) as the distortion function $d(\cdot)$, given as $d(\mathbf{s}_{i-1},\mathbf{s}_{i})=\|  \mathbf{s}_{i-1} \!-\! \mathbf{s}_{i}\|_2^2 /n$. The training objective can be then formulated as 
	\begin{equation}\label{sum_loss}
		L\left( \mathbf{s}_0, \bm{\theta},\bm{\phi} \right)=\mathbb{E}_{\mathbf{s}_0  \sim p_{\mathbf{s}}} \big[\sum_{i=1}^{t} 	\lambda^{t-i}L_i (\mathbf{s}_{i-1},\mathbf{s},h_i, \mathbf{n}_i)\big] ,
	\end{equation}
	where $\lambda$ is the introduced incremental weighting scheme to assign different weights to the loss function at different stages.
	Therefore, the parameters of the encoder and decoder can be obtained by solving
	\begin{equation}
		\left\{\bm{\theta}^*,\bm{\phi}^* \right\}=\arg \max_{\bm{\theta},\bm{\phi}} \; L\left( \mathbf{s}_0, \bm{\theta},\bm{\phi} \right).
	\end{equation}
	The detailed recursive training algorithm is summarized in Algorithm \ref{RecursiveAlgorithm}. 
	
	\begin{figure}[!htbp]
		\begin{centering}
			\includegraphics[width=0.38\textwidth]{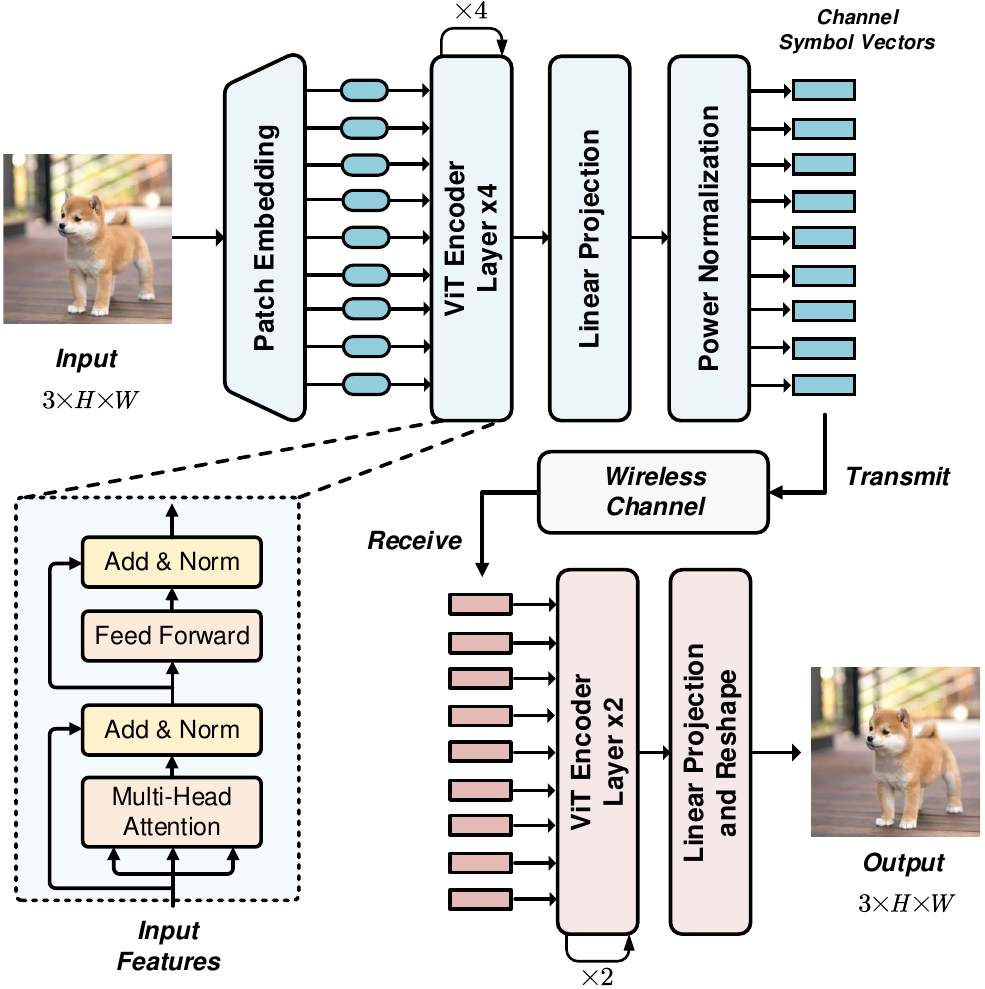}
			\par\end{centering}
		\caption{The proposed overall network architecture of ViTSC for image transmission.}
		\label{ModelArchitech}
	\end{figure}
	
	\subsection{Model Architecture}
	We utilize the architecture depicted in Fig. \ref{ModelArchitech} to construct the encoder and decoder for our semantic communication system, ViTSC. The encoder mainly comprises $4$ vision transformer (ViT) encoder layers, while the decoder consists of $2$ ViT encoder layers, known for their notable performance advancements in diverse image processing tasks. Each ViT encoder layer, as illustrated in Fig. \ref{ModelArchitech}, consists of a multi-head self-attention (MSA) module and a feed forward neural network layer. Additionally, GeLU activation and layer normalization operations are applied before the MSA and MLP modules.

	\section{Numerical Results} \label{S4}
	In this section, we present numerical results to demonstrate the performance of the proposed method. For communication settings, we consider both AWGN channel and Rayleigh fading channel. To ensure a fair comparison, we assume that all devices share the same wireless scenario, i.e., $h_i$ and $\mathbf{n}_i$ of different devices follow the same distributions, allowing for reasonable and unbiased performance evaluation.
	We implement the proposed system using the open-source tool, PyTorch. For optimization, we utilize the AdamW optimizer with an initial learning rate of $0.0005$, which decreases during the training epochs. The training batch size is set to $64$, and we set $\lambda$ to $0.9$.
	Furthermore, we train the semantic communication system using the CIFAR-10 dataset, which consists of $60,000$ images with a size of $3\times 32 \times 32$. In addition, we consider the cases that $\textrm{CBR}=1/6$ and $\textrm{CBR}=1/12$.

	The peak signal-to-noise ratio (PSNR) is selected as the performance metric, which is defined as 
	\begin{equation}
		\textrm{PSNR}=10 \log _{10}\left(\frac{\textrm{MAX}^2}{\textrm{MSE}}\right)=20 \log _{10}\left(\frac{\textrm{MAX}}{\sqrt{\textrm{MSE}}}\right),
	\end{equation}
	where $\textrm{MAX}$ is the maximum pixel value and equals $255$ for an image of $8$-bit representation, $\textrm{MSE}$ is calculated between the transmitted and received images. As for benchmarks, we consider the classic DeepJSCC proposed in \cite{Bouurt_TCCN}. In addition to DeepJSCC, we also consider the traditional separate coding scheme, i.e., BPG source coding combined with advanced LDPC channel coding. Besides, we adopt quadrature amplitude modulation (QAM) in traditional separate coding scheme. Furthermore, we use ViTSC+RT to denote training ViTSC with the proposed recursive training strategy.

	\begin{figure}[t]
		\centering
		\subfloat[]{\centering \scalebox{0.32}{\includegraphics{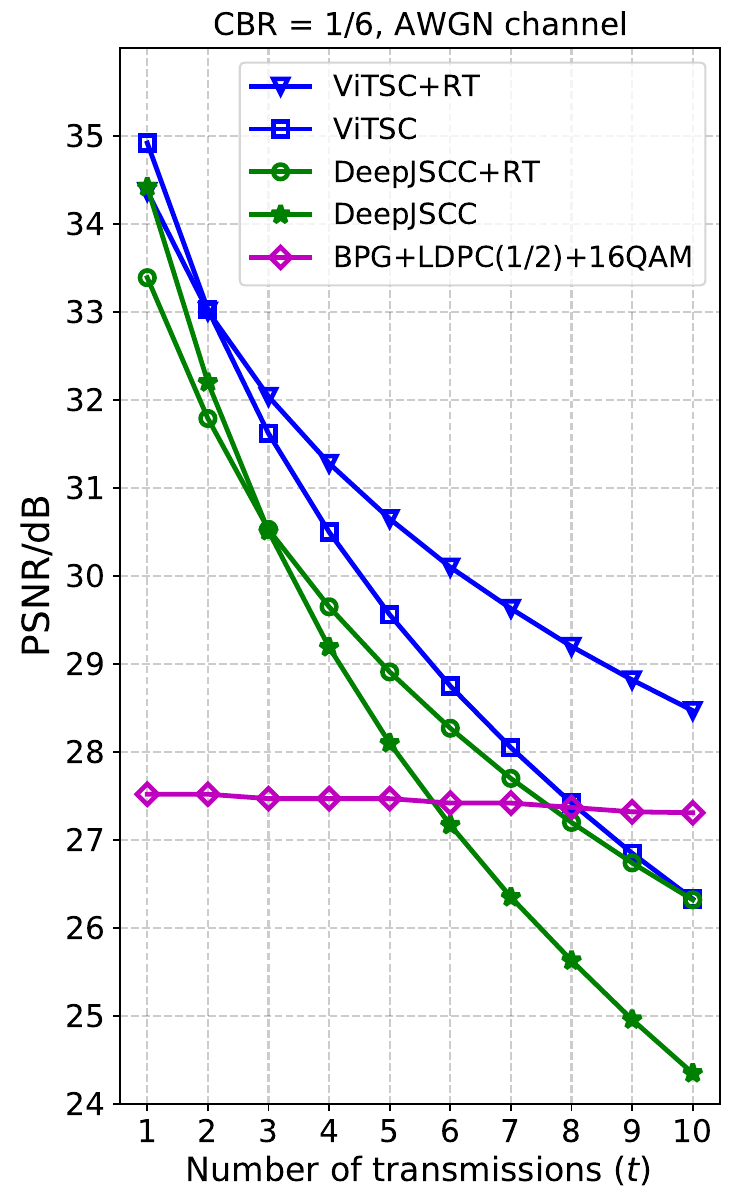}} }
		\subfloat[]{\centering \scalebox{0.32}{\includegraphics{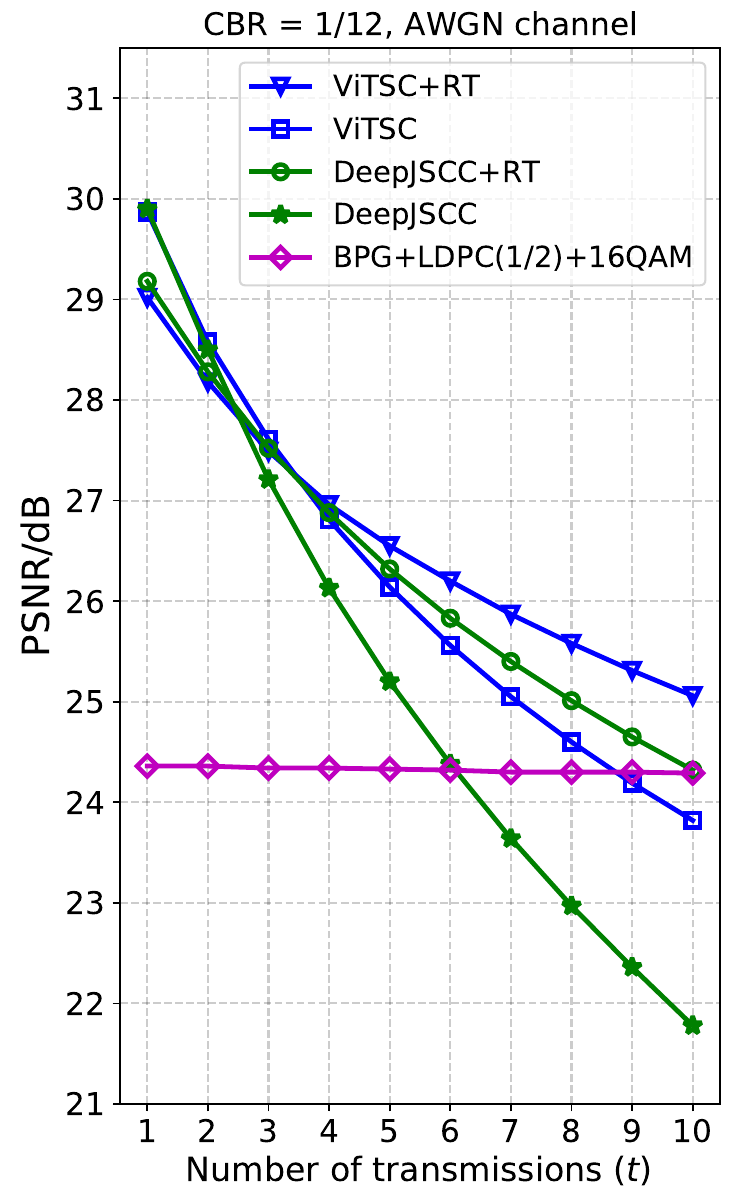}} }
		\caption{The PSNR performance versus the number of transmissions in AWGN channel at SNR$=18$ dB.}
		\label{SimFig1}
	\end{figure}

	\begin{figure}[t]
		\begin{centering}
			\includegraphics[width=0.34\textwidth]{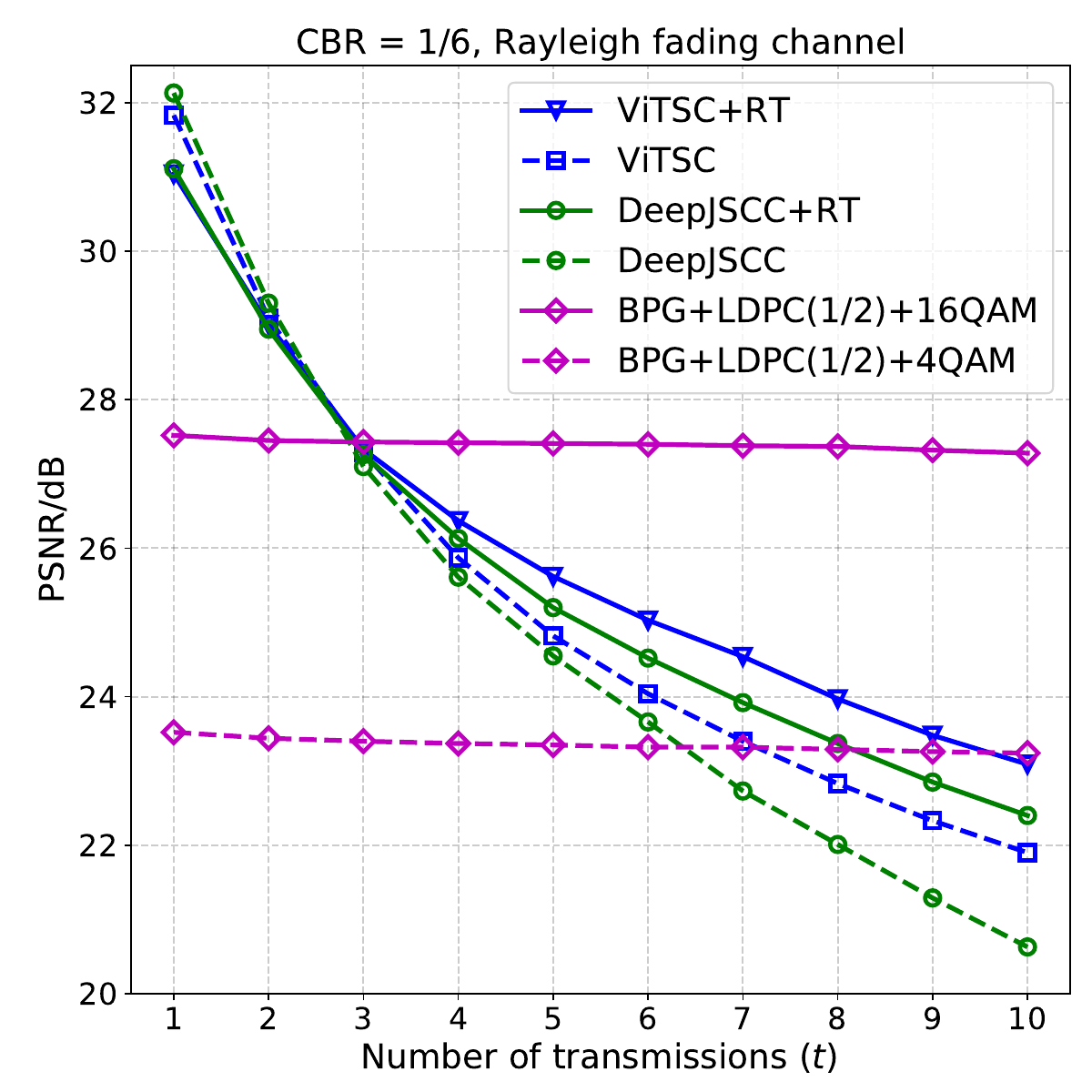}
			\par\end{centering}
		\caption{The PSNR performance versus the number of transmissions in Rayleigh fading channel at SNR$=18$ dB.}
		\label{SimFig2}
	\end{figure}

	\begin{figure}[t]
	\centering
	\subfloat[]{\centering \scalebox{0.32}{\includegraphics{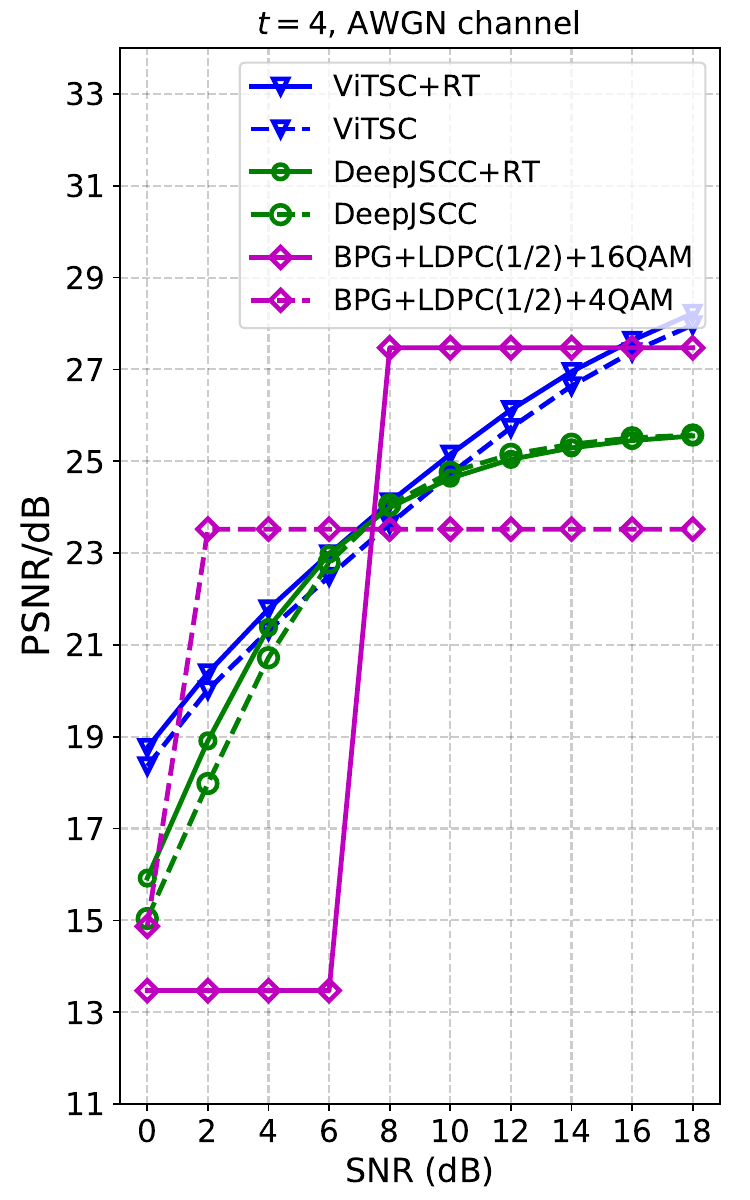}} }
	\subfloat[]{\centering \scalebox{0.32}{\includegraphics{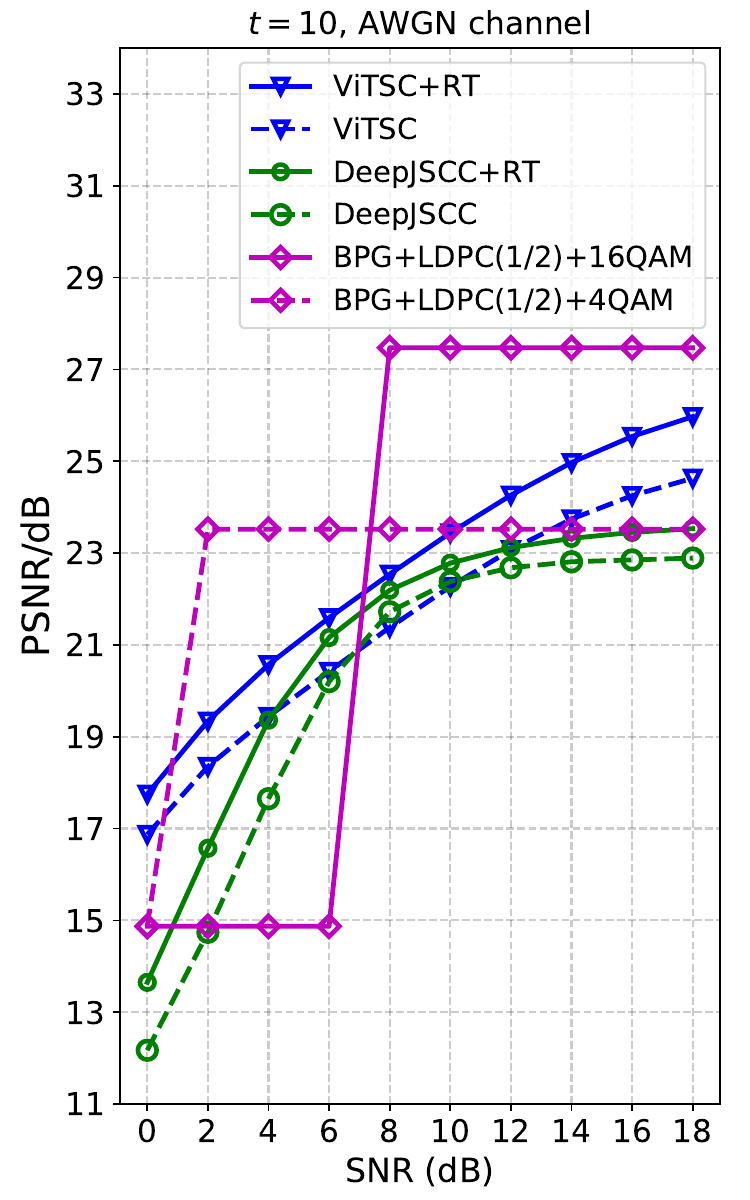}} }
	\caption{Performance comparison of proposed methods and benchmarks with respect to the channel SNR over an AWGN channel.}
	\label{SimFig3}
\end{figure}
	
	Fig. \ref{SimFig1} compares the PSNR performance of the proposed method and the benchmarks against the number of devices $t$, representing the number of transmissions, with CBR selected as $1/6$ and $1/12$. Additionally, all models are trained at SNR $= 18$ dB.
	From the figure, we observe that the achieved PSNR values of deep learning-based semantic communication systems, including ViTSC and DeepJSCC, decrease with the number of transmissions. This illustrates that deep learning-based methods suffer from the problem of distortion accumulation, leading to degraded image reconstruction quality as the number of transmissions increases.
	Furthermore, the proposed ViTSC outperforms DeepJSCC, showcasing its superiority. Moreover, by comparing ViTSC+RT with ViTSC, and DeepJSCC+RT with DeepJSCC, we observe that the proposed recursive training method significantly alleviates distortion accumulation. This is because recursive training helps ViTSC learn to deal with disturbed images at intermediate devices, showing the superiority of the proposed recursive training method. Moreover, we observe that the performance gain between ViTSC+RT and ViTSC increases as the number of devices $t$ increases.
	Additionally, we find that the PSNR performance of the separate coding scheme remains relatively stable with varying $t$. This stability arises because transmission errors are almost non-existent when transmission rate is kept below channel capacity. In this case, the image $\mathbf{s}_i$ received on device $u_i$ for $i\geq 2$ is almost identical to $\mathbf{s}_1$, thus there is almost no distortion accumulation. Furthermore, according to Fig. \ref{SimFig1}(a) and Fig. \ref{SimFig1}(b), we can conclude that the proposed recursive training method is effective under different CBR values.

	Fig. \ref{SimFig2} presents the PSNR performance under Rayleigh fading channel versus the number of devices $t$. All models are trained at SNR $= 18$ dB, and the channel state information is known at the receiver. It is shown that the achieved PSNR values of ViTSC and DeepJSCC decrease with the number of transmissions. Due to the fading effect, the problem of distortion accumulation is more critical than that in AWGN channel, making it harder to solve the problem. From the figure, we find that the recursive training method is still effective and can achieve relatively high performance gain with a large $t$, without performance degradation with a small $t$. This result emphasizes benefits of the proposed technique when communicating over different channels.
	
	Fig. \ref{SimFig3}(a) and Fig. \ref{SimFig3}(b) show the PSNR performance versus SNR under the AWGN channel. We present the PSNR performance of different schemes at the $4$-th and $10$-th devices. In addition, the ViTSC and DeepJSCC are trained under SNR sampled from $[0,2,4,6,8,10,12,14,16,18]$ dB. According to the results, the proposed ViTSC outperforms DeepJSCC and BPG+LDPC. Compared with DeepJSCC, ViTSC is able to provide a more graceful performance degradation as SNR decreases. Moreover, at low SNR regime, e.g., SNR $\leq8$ dB, the traditional BPG+LDPC suffers from cliff effect, since the channel condition is below the level anticipated by the LDPC code. In comparison, deep learning-based methods,  including ViTSC and DeepJSCC, can overcome the cliff effect. Nevertheless, due to the problem of distortion accumulation in multi-hop scenarios, the performance of deep learning-based methods with a larger $t$ gets worse. Moreover, it can be observed that the proposed methods can alleviate the distortion accumulation at different SNR regimes.
	
	\section{Conclusion} \label{S5}
	In this letter, we proposed a framework of multi-hop semantic communication for image transmission and employed a powerful ViT to implement the proposed semantic communication system. Moreover, since deep learning-based semantic communication systems are lossy transmission schemes, the distortion of images will accumulate with the number of transmissions. To deal with it, we developed a novel recursive training method with incremental weight.  Simulation results demonstrated that the proposed method can significantly alleviate distortion accumulation.

	\bibliographystyle{IEEEtran}
	\bibliography{IEEEabrv,Reference}
	
\end{document}